\documentclass[prl,aps,showpacs,twocolumn,unsortedaddress]{revtex4}
\usepackage{graphics,bm}
\usepackage{amssymb}
\usepackage{epsfig}
\usepackage{epsf}

\begin{document}

\title{Interlayer Transport in Bilayer Quantum Hall Systems}

\author{Enrico Rossi, Alvaro S. N\'u\~nez, and A.H. MacDonald}
\email{macd@physics.utexas.edu}
\homepage{http://www.ph.utexas.edu/~macdgrp}
\affiliation{Department of Physics, University of Texas at Austin, Austin TX 78712}

\date{\today}

\begin{abstract} 
Bilayer quantum Hall systems have a broken symmetry ground state at filling 
factor $\nu=1$ which can be viewed 
either as an excitonic superfluid or as a pseudospin ferromagnet.  We 
present a theory of inter-layer transport 
in quantum Hall bilayers that highlights remarkable similarities and critical differences between 
transport in Josephson junction and ferromagnetic metal spin-transfer devices.
Our theory is able to explain the size of the large but finite low bias 
interlayer conductance and the voltage width of this collective transport anomaly.   
\end{abstract} 

\maketitle

\noindent
{\em Introduction}--- Quantum Hall bilayers at total Landau level 
filling factor $\nu=1$ have ground states with 
spontaneous interlayer phase coherence.\cite{fertig}  These 
broken symmetry states can be regarded equivalently either as pseudospin 
ferromagnets or as excitonic superfluids.\cite{physica} One of the 
most spectacular experimental manifestation of this order is 
an enormous low-temperature enhancement\cite{eisenstein} 
of the interlayer tunneling conductance in samples with extremely 
small inter-layer tunneling amplitudes.  The differential tunnel  
conductance is large only at small bias voltages and reaches a finite 
maximum value that can be as large as 
$\sim 0.5 \, e^2/h$ at low temperatures. Although these important 
observations\cite{eisenstein,bilayerexpt} have inspired considerable interesting theoretical activity\cite{tunneltheory}
it has not been possible to account for their main qualitative features,
in particular for the voltage width of the anomaly, the finite value of the conductance
maximum, and the inverse relationship between these two quantities.
In this Letter we present a theory of the low-bias tunneling anomaly
which, in contrast to most previous theoretical work, predicts that the 
zero bias conductance is finite even in a perfect disorder free 
bilayer at temperature $T=0$, and accounts approximately for the width of the 
anomaly.  Our theory sees interlayer tunneling phenomena as partially
analogous to both tunneling across a Josephson junction\cite{JosephsonJunction} and spin-transfer 
phenomena in ferromagnetic metals.\cite{Slon,TransferExpt,TransferTheory}      
The key difference between these two examples of current driven order 
parameter manipulation is that the bias is applied by a superconducting 
condensate in the former case and by dissipative quasiparticles in the latter.
Tunneling in quantum Hall bilayers is an example of pseudospin transfer, 
with the feature particular to quantum Hall systems that transport quasiparticles are localized 
at the edge of the system when order is strong and the quantum Hall 
effect firmly established.

\noindent
{\em Josephson Junction and Spin Transfer Phenomenology}---
The semi-classical equations of motion for the collective dynamics of a 
current-biased Josephson Junction and a single-domain easy-plane ferromagnet
in an in-plane field 
are similar: 
\begin{eqnarray}
\hbar {\dot \phi} &=&  2e V  + \alpha_{\phi} {\dot N} \nonumber \\  
{\dot N} &=& - \frac{I_c}{2e} \sin(\phi) - \alpha_G \; \dot{\phi} + \frac{I_{bias}}{2e}
\label{JJdynamics}  
\end{eqnarray}  
and 
\begin{eqnarray}
\hbar {\dot \phi} &=&  \frac{K_z}{M_0} M_z   + \alpha_{\phi} {\dot M_z}  \nonumber \\  
{\dot M}_z &=& - \frac{H_x}{g \mu_B \hbar} \; M_0 \sin(\phi)  - \alpha_{\theta} M_0 \; {\dot \phi} + \frac{g_{ST} I}{e}. 
\label{STdynamics} 
\end{eqnarray}  
In Eq.(~\ref{JJdynamics}) $V$ is the voltage difference across the junction,
produced partly by the external circuit and partly capacitively by 
junction charging, $N$ is the number of 
Cooper pairs accumulated on one side of the junction, $\alpha_G$ 
is a dissipation coefficient due to thermally activated quasiparticle conductance 
and coupling to the external circuit and $\alpha_{\phi}$ is an additional
dissipation coefficient that is normally negligible (and often neglected),
and $I_{bias}$ is the bias current which represents the influence of the 
rest of the superconducting circuit on the junction region.  The 
{\em dc} Josephson effect occurs when the condition for quasiparticles to be in
equilibrium with the condensate, $I_c \sin(\phi) = I_{bias}$, is satisfied.  The current bias 
influences the microscopic self-energy of the quasiparticles so that the phase 
change across the junction $\phi$ is no longer zero when the quasiparticle 
density matrix has its equilibrium value. Current then flows across the junction 
without dissipation.

In Eq.(~\ref{STdynamics}),
$\alpha_{\phi}$ and $\alpha_{\theta}$ are the dissipation coefficients that appear in the 
Landau-Lifshitz-Gilbert (LLG) equations for ferromagnets,
$K_z$ is the magnetic anisotropy coefficient which is normally
dominated in thin film magnets by magnetostatic interactions 
(shape anisotropy), and $g_{ST} I/e$ is the spin-transfer torque.
The presence of this current bias term in the collective magnetization
equations of motion\cite{Slon,TransferTheory} can
be inferred from the approximate conservation of total spin in typical 
itinerant electron ferromagnets.  The version of the spin-torque LLG equations 
specified by Eq.(~\ref{STdynamics}) 
applies when the transport current incident on the magnetic nanoparticle of interest
has a spin orientation perpendicular to the easy plane, the circumstance relevant to tunneling in 
bilayer quantum Hall systems as we see below.  $g_{ST}$ is spin-transfer efficiency factor which is typically 
$\sim 1$ and depends on both the degree of polarization of the injected 
current and the rest of the circuit.  The spin-transfer torque term in the equation of motion,
like the bias term in the Josephson junction case, 
arises\cite{nuneztransfer} microscopically
from a change in quasiparticle self-energy;  
in the presence of the transport current, the quasiparticles are in equilibrium not when the
in-plane magnetization is aligned with the LLG equation effective magnetic field, but when its orientation in
the easy-plane is displaced from the field direction by an angle proportional
to the current.  The key difference between a current-biased Josephson junction and a 
current-biased nanomagnet is that the bias field experienced by the quasiparticles is applied 
in one case by the condensate, and in the other case by transport quasiparticles
that are held out of equilibrium by a finite bias voltage.  The current in spin-transport devices 
is always dissipative.  Unlike a Josephson junction, a finite voltage drop occurs across a 
nanomagnet when a spin-current bias is applied. 

\noindent
{\em Interlayer Transport in Quantum Hall Bilayers}---We now discuss interlayer 
tunneling in bilayer quantum Hall systems, starting with the ideal disorder free case. 
We use a pseudospin ferromagnet language in which electrons in the top layer have 
pseudospin up and electrons in the bottom layer have pseudospin down.  In the ordered state
quasiparticles in the bulk of a quantum Hall bilayer
\begin{figure}[!t]
\begin{center}
\includegraphics[width=0.48\textwidth]{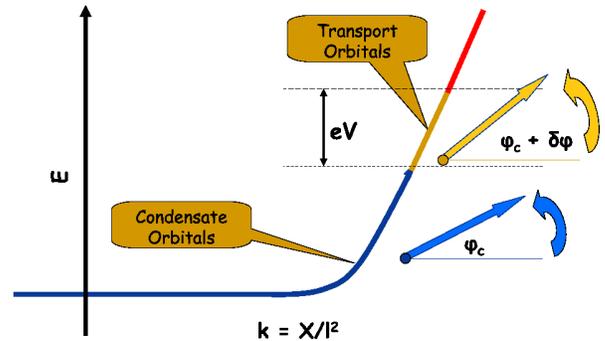}
\caption{Schematic illustration of pseudospin transfer torques in quantum Hall bilayers. The transport orbitals
on a quantum Hall plateau are extended along the edge of the system.  In the absence of disorder wavevector $k$ is a 
good quantum number and is proportional to cyclotron-orbit guiding center $X = k \ell^2$.  Each orbital makes 
a contribution to the pseudospin exchange field that is in the direction of its pseudospin orientation.  Unlike
equilibrium orbitals, transport orbitals do not align with their pseudospin fields, giving rise to a pseudospin
transfer torque.  Large interlayer conductance occurs when the pseudospin transfer torque can be canceled by a 
pseudospin torque due to the bare interlayer tunneling term in the Hamiltonian.}
\label{fig:one}
\end{center}
\end{figure} 
have an interlayer tunneling amplitude that includes a self-energy\cite{fertig,physica,
HartreeFockTheory} contribution which can be represented by an in-plane pseudospin effective 
magnetic field:
\begin{eqnarray}
\Delta_x(X) &=& \Delta_t + \frac{1}{L} \sum_{X'} n_{X'} F_D(X-X') \cos(\phi_{X'}) \nonumber \\
\Delta_y(X) &=& \frac{1}{L} \sum_{X'} n_{X'} F_D(X-X') \sin(\phi_{X'}) 
\label{EffectiveField}   
\end{eqnarray} 
In Eq.(~\ref{EffectiveField}), $\Delta_t$ is the single-particle tunneling amplitude which 
for the systems in question has an extremely small\cite{eisenstein} value $\sim 10^{-8} \; {\rm eV}$, $X = \ell^2 k$ labels
a guiding center state which is delocalized along the edge of the system (see Fig.[~\ref{fig:one}]), $n_{X'}$ is a guiding center 
state occupation number, $L$ is the length of the system along the edge, $\phi_{X'}$ is the planar 
pseudospin orientation (or equivalently
the phase difference between top and bottom layers) for guiding 
center $X'$ and $F_D(X-X')$ is\cite{HartreeFockTheory} the interlayer exchange integral 
between guiding centers $X$ and $X'$.  In the absence of a transport current, the quasiparticle 
pseudospin will align with the effective magnetic field for each guiding center:
\begin{equation}
\frac{\cos(\phi_X)}{\sin(\phi_X)} = \frac{\Delta_x(X)}{\Delta_y(X)}.
\label{EquMeanFieldEqs}
\end{equation} 
It is easy to verify that in this case the only self-consistent solution to Eqs.(~\ref{EffectiveField},~\ref{EquMeanFieldEqs}) is 
$\sin(\phi_X) \equiv 0$; the small single-particle tunneling amplitude selects the
phase difference between the two layers and the effective tunneling amplitude
is enormously enhanced by interactions ($\Delta_t \to \Delta_x(X)$).

When the system carries a current, the quasiparticle's Schroedinger equation must be 
solved with the scattering boundary condition that the current is incident from the high chemical potential 
contact.  Under these circumstances, its pseudospin orientation will {\em not} align \cite{nuneztransfer} with  
its effective magnetic field.  Since all electrons 
move along the edge at the magnetoplasmon velocity\cite{Wassermeier} $v_{emp} \sim 2 \times 10^{6} {\rm m}\,{\rm s}^{-1}$,
the quasiparticle Schroedinger equation may be mapped to that of a zero-dimensional spin $S=1/2$ particle in a time-dependent field.  
Differences between the disorder potentials in the two layers will give rise to a pseudospin effective field 
in the $\hat z$ direction which varies randomly along the edge.
Averaging along the edge, the rate of spin precession from up (top layer) to down (bottom layer) 
is proportional to the mean torque that acts on the planar spin.
We find that for the transport electrons 
\begin{equation}
\frac{\Delta_{QP}^{E}}{\hbar} \sin(\phi_{tr}-\phi_{c}) = \frac{g}{t_{transit}} = \frac{g v_{emp}}{L}.
\label{PseudoSpinTransferTorque} 
\end{equation}
In Eq.(~\ref{PseudoSpinTransferTorque}), $\phi_{c} = \tan^{-1}(\Delta_y/\Delta_x)$ is the orientation of the 
pseudospin effective field at the edge, $\Delta_{QP}^{E}$ is the magnitude of the 
exchange effective magnetic field at the edge of the system ($\Delta_{QP}^{E} = L^{-1} \sum_{X'<X} F_D(X-X')$ up to negligible
corrections from $\Delta_{t}$ and is half the bulk quasiparticle gap), $g$ is the probability that an electron injected in the 
top layer will make its way to the bottom layer, and $t_{transit} = L/v_{emp}$ is the time required 
for an edge electron to transit the sample.  Eq.(~\ref{PseudoSpinTransferTorque}) is most easily understood
in a reference frame that moves along the edge (at the magnetoplasmon velocity) with the 
transport electrons; from this point of view it merely asserts that a spin-torque due to misalignment 
between the exchange field and the in-plane pseudospin-orientation is necessary to produce
the required precession from top layer (pseudospin up) to bottom layer (pseudospin down).
We refer to $\phi_{c}$ below as the condensate pseudospin orientation.
Since $g < 1$, $\Delta_{QP} \sim 10^{-4} {\rm eV}$ \cite{HartreeFockTheory} and $L \sim 10^{-2}{\rm cm}$ in typical samples, it 
follows that $\delta {\phi} \equiv \phi_{tr}-\phi_{c}$ is small and therefore that $\sin(\delta \phi) 
\approx \delta \phi$.

We propose that the tunneling anomaly in quantum Hall bilayers occurs when it is 
possible to achieve a self-consistent solution of the mean-field equations in the presence of current 
induced pseudospin-torques.  Combining Eq.(~\ref{EffectiveField}) 
and Eq.(~\ref{PseudoSpinTransferTorque}) we obtain 
\begin{eqnarray}
\Delta_x(X_e) &=& \Delta_t + \Delta_{QP}^{E} \cos(\phi_c) + \frac{F_D(0) N_{tr}}{L} \cos(\phi_c +\delta \phi) \nonumber \\
\Delta_y(X_e) &=& \Delta_{QP}^{E} \sin(\phi_c) + \frac{F_D(0) N_{tr}}{L} \sin(\phi_c +\delta \phi).
\label{deflectedpseudospin}   
\end{eqnarray} 
In this equation $N_{tr} = L eV/(h v_{emp})$ is the number of edge states in the narrow transport window
with energy width $e V$.  (See Fig.[~\ref{fig:one}].) The bias voltage is assumed to be small enough that the 
guiding center width of the transport window is much smaller than the range\cite{HartreeFockTheory} 
($\sim \ell$ where $\ell$ is the magnetic length) of the $F_D(X-X')$ 
exchange integral.  The terms proportional to $\Delta_{QP}^{E}$ in Eq.(~\ref{deflectedpseudospin}) are edge
self-energies in the absence of transport currents and the terms proportional to $F_D(0)$ are the 
pseudospin torque contributions.  Eqs.(~\ref{deflectedpseudospin}) neglect the variation of $\phi_c$ 
from its maximally deflected value at the edge to the its value deep in the bulk ($\phi_c=0$) since these 
occur on a length scale $(\ell \sqrt{\Delta_{QP}^{E}/\Delta_t}\;)$ much longer than the 
range of $F_D(X-X')$.  Solving Eqs.(~\ref{deflectedpseudospin}) we find that 
\begin{equation} 
\Delta_t \sin(\phi_c) = \frac{F_D(0) N_{tr}}{L} \delta \phi
\end{equation} 
and hence that the maximum bias voltage for which a time-independent solution 
of the mean-field equations exists is 
\begin{equation}
eV^* = \frac{ \Delta_t v_{emp} h}{F_D(0) \delta \phi} = \frac{2 \pi \Delta_t \Delta_{QP}^{E} L}{g F_D(0)}.
\label{voltagewidth}  
\end{equation} 
Since\cite{HartreeFockTheory} $F_D(0) \sim e^2/(\epsilon) \sim 10^{-2} {\rm eV} \ell$, it follows that 
the width of the low-bias tunneling anomaly when the 
quantum Hall effect is most strongly developed
(at the lowest temperatures) 
should be $\sim 10^{-6} \,{\rm eV}$, consistent with experiment. 

\noindent
\begin{figure}[!t]
\begin{center}
\includegraphics[width=0.48\textwidth]{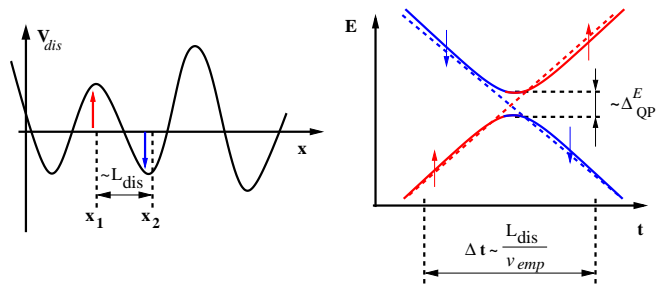}
\caption{Schematic illustration of the suppression of interlayer tunneling by 
         disorder.  Differences between the random potentials in the two layers give rise to
substantial pseudospin fields in the $\hat z$ direction $V_{dis}$ which are not effectively 
screened at the edge of the system.  Along much of the edge, $V_{dis}$ is larger than the 
in-plane coherence induced exchange field $\Delta_{QP}^{E}$.  When the disorder potential
difference changes sign, quasiparticles will cross between layers if they follow the 
adiabatic path.  Because of the high velocity of edge excitations, quasiparticles are likely
to Landau-Zener tunnel to the higher energy state and remain in the same layer.}
\label{fig:two}
\end{center}
\end{figure} 
{\em Landau-Zener Tunneling}--Having established our basic picture, we now comment on the 
surprisingly rapid decrease of the 
interlayer conductance with increasing temperature, and on the extreme sample quality required to approach
the highest zero-bias conductance values.
When a time independent solution of the mean-field equations is possible,
the quasiparticle conductance can be evaluated using the Landauer-Buttiker\cite{Landauer} scattering
theory picture of transport which predicts conductance $g \, e^2/h$.  The maximum possible value $g=1/2$
applies when an electron injected in one 
layer has a 50-50 chance of being found in either layer at later times.  In the absence of disorder 
$g$ should approach $0.5$ when the pseudospin precession length $L_{pr} \equiv v_{emp}\, \hbar /\Delta_{QP}^{E}$
is less than $L$, {\em i.e.} when $\Delta_{QP}^{E}$ is larger than $\sim 10^{-5} eV$, a condition that 
we expect to be satisfied quickly once the ordered state is entered.  Instead $g$ is almost always 
substantially smaller than $1$ in experiment.  Interlayer tunneling that is strongly suppressed 
compared to naive expectations has been seen\cite{Sakaki,Yoshioka} previously in bilayers with large 
purely single-particle tunneling amplitudes.  In both cases, we ascribe the behavior to the combined effects 
of disorder and large edge magnetoplasmon velocities.  

As illustrated in Fig.[~\ref{fig:two}] electrons traveling along the edge at
velocity $v_{emp}$ see differences between the random disorder potentials
in the two layers as $\hat z$ direction pseudospin magnetic fields.  
The typical rate at which the pseudospin field varies is 
$v_{emp} V_{dis}/L_{dis}$ where $V_{dis}$ and $L_{dis}$ are the typical size and correlation 
length of the potential difference between the layers.  (In the bulk of the two-dimensional electron system
these random pseudospin field fluctuations are screened\cite{fertig,HartreeFockTheory} by tilting the
pseudospin slightly out of the easy plane.) 
From floating gate screened disorder potential measurements in the quantum Hall regime\cite{Yacobi} 
it follows that $L_{dis} \sim 10^{-6} {\rm m}$ so that $\sim L/L_{dis} \sim 100$.
Avoided crossings of adiabatic pseudospin energy 
levels occur when the disorder potential difference changes sign. 
Quasiparticles that follow the 
adiabatic evolution path will transfer between layers at each avoided crossing. 
We estimate $g$ as the product of the number of avoided crossings (disorder potential difference sign
changes) and the probability of transfering between layers at an individual crossing
\cite{LandauZener_ref}:
\begin{eqnarray} 
g &=& (L/L_{dis}) (1 - \exp(-2(\Delta_{QP}^{E})^2 L_{dis}/\hbar v_{emp} V_{dis})) \nonumber \\
 &\sim&  2(\Delta_{QP}^{E})^2 L/\hbar v_{emp} V_{dis}.
\label{LandauZener} 
\end{eqnarray}
The strength of the bare disorder potential $V_{dis}$ in typical high-mobility samples
can be estimated by floating gate measurements\cite{Yacobi}; we find that 
$V_{dis}$ is comparable to the zero-field Fermi energy $\sim 10^{-3} {\rm eV}$.
It follows that the argument of the exponential function in the Landau-Zener 
tunneling formula is small even for the $T=0$ values of $\Delta_{QP}^{E}$, 
consistent with experiment.  This rough estimate of the typical value of $V_{dis}$ 
is consistent with the observations that $g$ approaches $1/2$ only at the lowest 
temperatures and justifies the small argument expansion used in the final form for 
the right hand side of Eq.(~\ref{LandauZener}). 

Eq.(~\ref{PseudoSpinTransferTorque}) can be applied consistently with Eq.(~\ref{LandauZener})
because of the long range of the Coulombic edge exchange interaction which 
can average over a number of potential inter-layer tunneling sites. 
Combining Eq.(~\ref{LandauZener}) and Eq.(~\ref{voltagewidth}) we find that 
\begin{eqnarray} 
eV^{*} = \frac{\Delta_t}{\Delta_{QP}^{E}} \frac{ h V_{dis} v_{emp}}{F_D(0)}. 
\label{WidthLandauZener}
\end{eqnarray} 
Eq.(~\ref{LandauZener}) for the zero-bias inter-layer tunnel conductance and Eq.(~\ref{WidthLandauZener}) 
for the voltage width of the low-bias anomaly are the main predictions of this paper.  
In our theory, the temperature dependence of the transport anomaly follows from 
thermal fluctuations in the condensate phase which 
reduce the order parameter and $\Delta_{QP}^{E}$.\cite{Benfatto,HartreeFockTheory}
The increase\cite{eisenstein} of $eV^*$ by a factor of approximately 40 between 
20 mK and 0.3 K is consistent with the size of suppression that is expected, 
although the detailed behavior is certainly disorder-dependent and sample 
specific.  The decrease\cite{eisenstein} in zero-bias conductance by a factor of 2000 
over the same temperature interval is then consistent with the predictions of Eq.(~\ref{LandauZener}).
Our theory also accounts qualitatively for in-plane field $B_{\parallel}$ dependence of the anomaly
which is marked\cite{eisenstein} by a strong decrease in conductance with little change in 
voltage width.  This behavior is predicted by our theory since $\Delta_{QP}^{E}$ and 
$\Delta_t$ have similar field dependence, both dropping\cite{HartreeFockTheory,JunHu} by a factor $\sim 1$ when
$B_{\parallel}/B_{perp} \sim d/\ell$. 

The authors gratefully acknowledge helpful interactions with Anton Burkov, Michel Devoret, Jim Eisenstein, 
Yogesh Joglekar, Mindy Kellogg, and Ian Spielman.  This work was supported in part by the Welch Foundation.      



\end{document}